\documentclass[twocolumn,showpacs,amsmath,amssymb,prl]{revtex4}
\usepackage{dcolumn}
\usepackage{epsfig}
\usepackage[latin1]{inputenc} 
\usepackage{bm}

\begin{document}

%
%
%
\title{Rigidity and intermediate phases in glasses driven by speciation} 
\author{Matthieu Micoulaut}
\affiliation{Laboratoire de Physique Théorique de la Matière Condensée,
Université Pierre et Marie Curie,  Boite 121\\ 4, Place Jussieu, 75252
Paris Cedex 05, France\\}

\date{\today}
\begin{abstract}
The rigid to floppy  transitions and the associated intermediate phase
in glasses are  studied in the case  where the local  structure is not
fully  determined  from the  macroscopic  concentration.  The approach
uses  size increasing cluster  approximations and  constraint counting
algorithms.   It is  shown  that the  location  and  the width of  the
intermediate phase  and the  corresponding structural,  mechanical and
energetical  properties of the network    depend crucially on the  way
local structures are selected at a given concentration. The broadening
of the intermediate phase  is obtained for  networks combining a large
amount of flexible  local structural units  and a high rate of  medium
range order.
\end{abstract}
\maketitle
Concepts from mean-field rigidity  in networks found their origin from
Lagrangian  constraint  counting in mechanics  \cite{Maxwell} and have
been  applied with great success in  glass science for several decades
\cite{rigidity}. Bonds in a glass network  can indeed be considered as
constraints    arising from    interatomic   stretching   and  bending
forces. The connectivity or cross-link density (best quantified by the
network  mean  coordination number  $\bar r$)  plays   therefore a key
role. In highly cross-linked networks  where $\bar r$ is large,  there
are more constraints than degrees  of freedom per  atom on average and
the structure is stressed rigid  (hyperstatic or overconstrained).  At
low connectivity, one  has  a flexible (hypostatic   or underconstrained)
structure that contains   more  degrees of freedom  than  constraints.
Thorpe \cite{Thorpe83} analyzed the vibrational behaviour of such kind
of  networks and  identified    a  mean-field (MF)  floppy  to   rigid
transition when the   mean  coordination number equals   $\bar  r=\bar
r_c=2.38$,  a  result that   agrees with global   (Maxwell) constraint
counting  as   enunciated   by  Phillips  \cite{JNCS1979}   from   the
enumeration $n_c$ of bond stretching and bond bending forces.
\par
The underlying nature   of this peculiar transition has  been
deeply  reinvestigated    recently 
because  two transitions at  $\bar  r_{c(1)}$ and  $\bar  r_{c(2)}$
have    been   found
\cite{Bool0} experimentally  in a  variety of glasses. These
 define  an
intervening region  (or intermediate phase,  (IP))  between the floppy
and  the  stressed  rigid  phase.  In the  IP,  glasses   display some
remarkable properties such as absence of ageing
\cite{JPhys2005} or  stress \cite{PRB2005}, selection of isostatically
rigid  local    structures   \cite{Bool0}   or    weak   birefringence
\cite{optical}.  The  two   boundaries have  been   characterized from
numerical calculations  \cite{JNCS2000}    and      cluster   analysis
\cite{PRB2003} on self-organized networks   and identified as being  a
rigidity transition  at low $\bar r$  and a stress transition  at high
$\bar  r$.  In the mean-field  approach   or in random  networks where
self-organization does not take place, both  transitions coalesce into a
single     one.   Moreover, links   between  IP    and protein folding
\cite{protein}, high-temperature superconductors
\cite{HTSC} or 
computational phase  transitions \cite{comput} have been stressed that
go  much beyond  simple  analogies.  The  understanding  of the IP  is
therefore  of  general  interest.  It  has   become  clear that stress
avoidance  in the network is responsable  for the width $\Delta \bar r=\bar
r_{c(2)}-\bar r_{c(1)}$ and the location of the intermediate phase, an
idea  that has gained  some strength from energetical adaptation in a
simple random bond model
\cite{Bishop} for the rigidity transition or suppressed nucleation of 
rigidity during a fluid-solid transition \cite{Huerta}. Mousseau and co-workers 
\cite{Mousseau} have also
shown  recently  that self-organization with  equilibration on diluted
triangular lattices would lead to an intermediate phase.
\par
However, the recent discovery of an IP in  more complex glass systems such as
silicates  \cite{Yann}  raises a    new  challenging issue.    On the
experimental side, most of the results have been obtained up to now on
simple  network  glasses (e.g.  the archetypal  $Ge_xSe_{1-x}$), where
$\bar r$ can be directly  related to the concentration ($\bar r=2+2x$)
of the species involved \cite{PRL97}. This  happens to be not the case
any more in  multicomponent  systems  such  as (even simple)    binary
glasses (a network former, e.g. $SiO_2$  and a modifier, e.g. $Li_2O$)
where a non-trivial speciation can appear  depending on the nature of
the  cation or the atoms involved.  This contributes to  $\bar r$ in a
non-linear fashion and application of constraint counting algorithms 
becomes more difficult.
\begin{figure}
\begin{center}
\epsfig{width=0.85\linewidth,figure=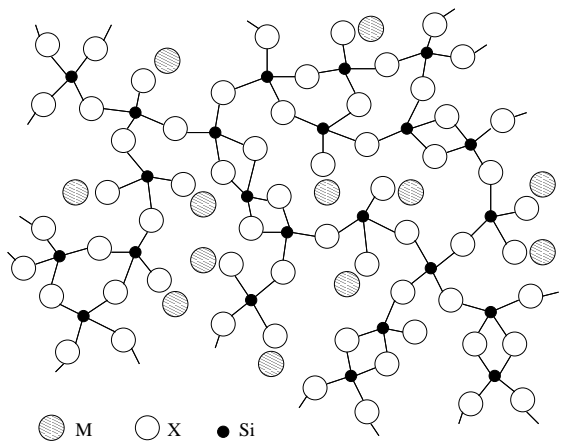}
\end{center}
\caption{\label{network} An example of a network made of 23 tetrahedra 
with   4  $Q^2$, 4  $Q^3$  and 15  $Q^4$  units,  corresponding to the
modifier  concentration   $x=0.207$.   The  fraction   of edge-sharing
tetrahedra is $\eta=0.32$.}
\end{figure}
We describe in this Letter how speciation affects onset of rigidity and the
intermediate  phase.  A simple model to highlight the
effect of   the   speciation is solved, and combined with cluster-constraint
calculations   applied  to  silicate  or  thiosilicates   of  the form
$(1-x)SiX_2-xM_2X$ with ($X=O,S,Se$) and ($M=Li,Na,K$) which are known
\cite{Science}  to   display  a   mean-field  rigidity transition   at
$x=0.20$.  The presence  of an intermediate  phase is demonstrated and
its   structural,      mechanical      and   energetical    properties
characterized. It appears that   the selection of more  flexible local
structural units with addition of a  modifier leads to a broadening of
the IP, independently of the degree  of medium range order. The latter
contributes however to the increase of $\Delta\bar r$ as well. On the other
hand,  the possibility of the system  to adapt the speciation in order
to lower the constraint free energy in  the stressed rigid phase leads
to a situation that ressembles very much to sodium silicates. Taken
together,   these results provide  new   benchmarks to  study IP's  in
multicomponent oxide or chalcogenide glass  systems such as fast ionic
amorphous  superconductors  where the  speciation and henceforth the elastic nature
of the network crucially determine
physical and electric transport properties.
\par
In binary   sodium  silicates (X=O,  M=Na),  speciation depends weakly
\cite{Mysen}  on the  nature  of the modifier  cation  M which creates
almost  only $Q^{n=3}$  units at low   $x$ so that  the probability of
finding the latter   is $R=2x/(1-x)$.  Here,  the superscript  "$n=3$"
denotes the  number   of bridging oxygens   on  a $SiO_{4-n/2}M_{4-n}$
tetrahedron (a $Q^n$  unit) that connects  to the rest of the  network
(Fig. \ref{network}). This means that the chemical reaction \cite{reaction}:
\begin{eqnarray}
\label{chemic}
 2Q^3\rightleftharpoons Q^4+Q^2
\end{eqnarray}
  involving the species $Q^n$ is desequilibrated on  the left side.  A
  radically different  situation  is   encountered  in systems    with
  modifier cations of smaller  sizes (M=Li) or in  thiosilicates (X=S)
  and seleniosilicates (X=Se)  \cite{Pradel2}. Certain of  these glass
  networks can be indeed made out of $Q^4$  and $Q^2$ species only or,
  at least, of a mixture of all  $Q^n$'s.  Noteworthy is the fact that
  Maxwell  constraint  counting  \cite{JNCS1979}  does not distinguish
  between the aforementioned   systems, although the location of   the
  rigidity and stress transitions should be obviously changed.
\par
Size   increasing     cluster  approximations  (SICA)   can    be used
\cite{PRB2003} to infer  the effect of speciation  on the location and
the   width of  the  intermediate phase.  We  consider a  network of N
tetrahedra $Q^4$, $Q^3$    and  $Q^2$ (see Fig.  \ref{network})   with
respective probabilities $p_4^{(1)}$, $p_3^{(1)}$ and $p_2^{(1)}$. The
behaviour of the $p_i^{(1)}$'s   with  modifier concentration $x$   can   be
determined      from              the      normalization     condition
$p_4^{(1)}+p_3^{(1)}+p_2^{(1)}=1$, the  charge        conservation law
\cite{Bray} $R=p_3^{(1)}+2p_2^{(1)}$, and finally the
definition  of  the  equilibrium  constant \cite{Ke}   of the chemical
reaction            (\ref{chemic})            given                by:
$K_e=p_4^{(1)}p_2^{(1)}/p_3^{(1)}p_3^{(1)}$.  For instance, in lithium
silicates  \cite{Schramm}   $K_e$  is   of the    order of   $0.3$  at
$x=0.17$.  With these  equations, the  speciation  is fully determined
with respect to $x$ and given by:
\begin{eqnarray}
\label{1}
p_3^{(1)}={\frac {R(2-R)}{1+\sqrt{(1-R)^2+4K_eR(2-R)}}}
\end{eqnarray}
out  of    which     is  obtained
$p_2^{(1)}=(R-p_3^{(1)})/2$                                        and
$p_4^{(1)}=1-p_3^{(1)}-p_2^{(1)}$.   Starting  from  this  short-range
order distribution  $p_i^{(1)}$ (the basic  SICA  units at the initial
step $l=1$ which will serve as building blocks), one constructs the 12
possible   structural arrangements   of  two  basic  units  ($l=2$),
i.e. $Q^4-Q^4$,   $Q^4-Q^3$, $Q^4-Q^2$, $Q^3-Q^3$, etc.   Three energy
gains,  $E_{stress}$,  $E_{iso}$  and $E_{flex}$ with  corresponding
Boltzmann  factors  $e_i=\exp[-E_i/k_BT]$,  are  defined following the
mechanical  nature  of    the   created   cluster  (stressed    rigid,
isostatically rigid and flexible).  The   probabilities of the   created
clusters       ($l=2$)   are    then     given    by    $p_{kj}^{(2)}\propto
W_{kj}p_k^{(1)}p_j^{(1)}e_i$ ($i=stress,iso,flex$) where $W_{kj}$ is
a statistical factor taking into account the number of equivalent ways
to   connect two ($l=1$)  units  together.  For   instance, there  are
$W_{44}=72$   different ways to    connect  two  $Q^4$ tetrahedra   by
edges.  One should also note that  there are only  two stressed rigid
clusters created (corner  and edge-sharing $Q^4-Q^4$ connections)  and
one isostatically rigid cluster ($n_c=3.0$, a corner-sharing $Q^4-Q^3$
connection).
\begin{figure}
\begin{center}
\epsfig{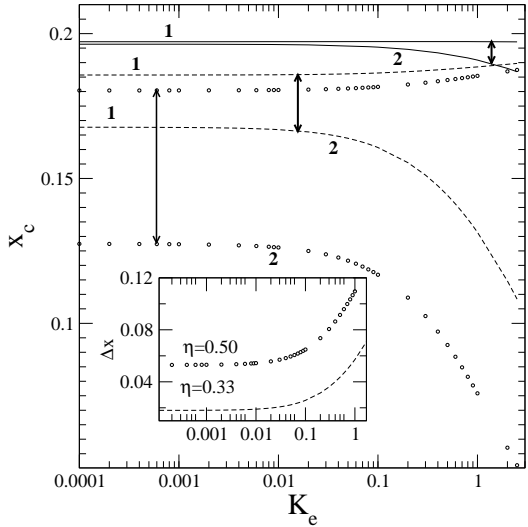}
\end{center}
\caption{\label{xc} Location of the rigidity (label 1) and the stress 
transition (label  2) as a function of  the equilibrium constant $K_e$
for  three   fractions of edge-sharing  tetrahedra  in  the base glass
$\eta=0.04$    (solid  line),  $\eta=0.33$    (broken   line) and   $\eta=0.50$
(dots).  Vertical  arrows  serve  to   identify the  width $\Delta x$  of  the
intermediate    phase at  a  given   $K_e$.    The insert shows  the
corresponding width of the intermediate phase.}
\end{figure}
\begin{figure}
\begin{center}
\epsfig{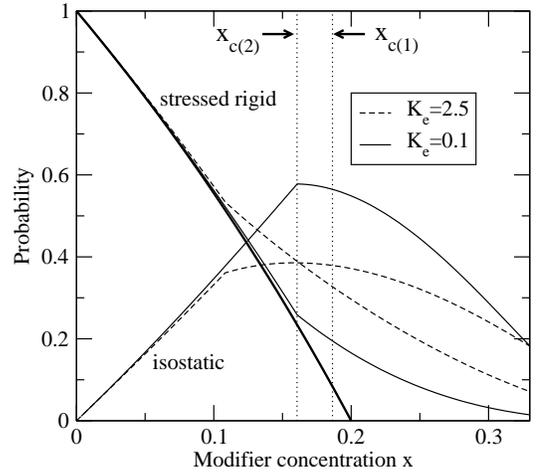}
\end{center}
\caption{\label{probclust} Probability of finding stressed rigid and 
isostatically  rigid  clusters with  respect  to   the modifier
concentration  $x$    for  two     different   equilibrium  constants:
$K_e=0.1$  (solid line) and $K_e=2.5$  (broken line). Note that the
stressed rigid  cluster probability extrapolates  to $x_c=0.20$ in the
MF description (bold line).  The dotted vertical lines serve to define
the two transitions and the intermediate phase for $K_e=0.1$.  Here
$\eta=0.33$.}
\end{figure}
Maxwell constraint counting \cite{Thorpe83} is then applied on the set
of ($l=2$) clusters that leads to the number of floppy modes
of the network given by:
\begin{eqnarray}
\label{f}
f^{(2)}=3-n_c^{(2)}=3-{\frac
{\sum_{k,j}n_{c(kj)}p_{kj}^{(2)}}{\sum_{k,j}N_{kj}p_{kj}^{(2)}}}
\end{eqnarray}
where  $n_{c(kj)}$ and $N_{kj}$  are     respectively the number  of
mechanical constraints  and the  number  of atoms  of the cluster with
probability $p_{kj}^{(2)}$.  Once this  is set and starting from a flexible (floppy)
network  where  stressed rigid   dendritic $Q^4-Q^4$   connections are
absent  and decreasing $x$, one can investigate at which concentration
$x_{c(1)}$ the  network will have  a vanishing  of $f^{(2)}$ (rigidity
transition) and at which concentration $x_{c(2)}$ the network will not
be    able    to avoid   stressed    rigid   dendritic  clusters (i.e.
corner-sharing $Q^4$) any more (stress transition). These corner-sharing $Q^4$'s
contribute to percolation of stressed rigidity. The calculation is
performed  for a given  amount of medium  range order characterized by
the fraction  $\eta$  of  edge-sharing  tetrahedra  in   the base   glass
($x=0$). This furtermore suggests that  the strain can be concentrated
in small rings and structures.
\par
Fig. \ref{xc} shows the results of the construction. Large equilibrium
constant  $K_e$ (corresponding to a  $Q^2$-rich  glass) will induce  a
large width  for the  IP.  However, one   observes  that $K_e$  mostly
affects  the location $x_{c(2)}$  of the stress transition whereas the
location  of    the  rigidity transition $x_{c(1)}$     remains almost
constant, as  already  signaled \cite{PRB2003}  for  a network  glass.
Noteworthy is    also the shift  from  the  MF rigidity  transition at
$x_c=0.20$ to  lower  $x$  that  arises  from the  presence  of weakly
stressed  rigid edge-sharing tetrahedra  (EST,  $n_c=3.33$  per atom).
With a weaker rigidity  due to the presence  of these EST, part of the
strain is captured in the EST, and  onset of flexibility can happen at
lower  modifier concentration   $x$.   As for IV-VI  network   glasses
\cite{PRB2003}, the width $\Delta  x=x_{c(2)}-x_{c(1)}$ of the IP increases
with the fraction $\eta$ of  EST due to the   shift of the location  of  the
stress transition  (Fig.  \ref{xc}). The change  in  speciation from a
$Q^3$-rich to a $Q^2$-rich glass contributes  however to an additional
broadening of the IP.  The trend with  $K_e$ observed in Fig. \ref{xc}
can  be further characterized from the  computation of the probability
of stressed rigid  and isostatically rigid  clusters using SICA.  When
the chemical  reaction (\ref{chemic})  is desequilibrated on  the left
side (low $K_e$, i.e.   a  $Q^3$-rich glass), each modifier   molecule
will create mostly two flexible $Q^3$ units ($n_c=2.55$ per atom) that
serve to accumulate isostatically  rigid subregions of the network, as
$Q^4-Q^3$ connections are likely to  appear. These are maximum at  the
stress transition (solid   line, Fig.  \ref{probclust}),  consistently
with numerical simulations
\cite{JNCS2000}. On  the other hand, a  higher value  of $K_e$
leads to the growth of even more flexible units ($Q^2$'s, $n_c=2.0$ per
atom) to the expense of $Q^3$'s, and will produce a stress transition at
lower $x$   (broken line,  Fig. \ref{probclust}).  Indeed,  increasing
$K_e$ at fixed $x$ decreases the network  mean coordination number and
favours  flexible $Q^4-Q^2$ instead  of  isostatically rigid $Q^4-Q^3$
bondings.  As a result,  with  growing concentration $x$,  the network
will lose stress earlier and will display a lower isostaticity in the
IP. Thus $x_{c(2)}$ is shifted to lower $x$.
\par
The constraint-related free eneregy is now considered, following the 
approach initially reported by Naumis \cite{Naumis}. 
 The  free energy of the system is given by:
\begin{eqnarray}
\label{free}
{\cal F}_{(2)}(x,K_e)=-f^{(2)}+k_BT\sum_{k,j}p_{kj}^{(2)}\ln p_{kj}^{(2)}
\end{eqnarray}
where $-f^{(2)}$ is the stress energy equal to the number of redundant
constraints, i.e. additional constraints  that  cannot be balanced  by the
degrees of freedom, and which vanish for $x>x_{c(1)}$.   
\begin{figure}
\begin{center}
\epsfig{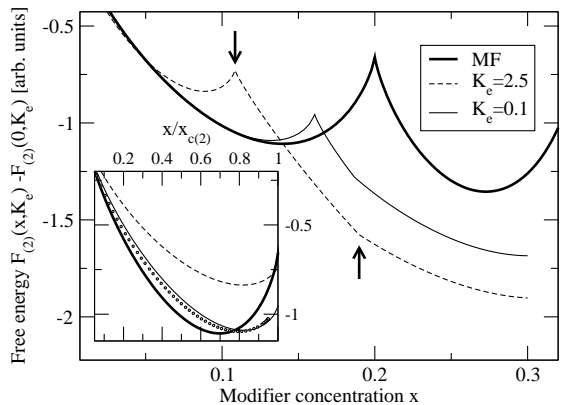}
\end{center}
\caption{\label{energy} Free energy ${\cal F}_{(2)}(x,K_e)-{\cal F}_{(2)}(0,K_e)$ 
of the system as  a function of  the modifier concentration $x$ in the
mean-field   case (bold solid line),  and   for two  different   equilibrium  constants:
$K_e=0.1$  (solid,    line)  and $K_e=2.5$   (broken  line),  both for
$\eta=0.33$. The arrows  indicate the stress  and rigidity transition for
$K_e=2.5$.   The insert shows the same  quantity  with $x$ rescaled to
its stress transition composition  $x_{c(2)}$, together with  the free
energy (dots) minimized by $K_e$. In  all, $k_BT=1$.}
\end{figure}
Figure \ref{energy} shows  that the stress  transition at $x=x_{c(2)}$
is first order for any $K_e$.  However, with respect to the mean-field
case (bold solid  line) where  $x_{c(1))}=x_{c(2)}$,  the jump of  the
first derivative $\partial{\cal F}_{(2)}(x,K_e)/\partial x$ at $x=x_{c(2)}$ decreases
with growing $K_e$.  In the MF  case, this jump  is  equal to $75.22$,
whereas it  is only  $45.30$  for $K_e=2.5$.  This suggests  that the
transition broadens  when the  equilibrium (\ref{chemic}) displaces to
the right side, leading to a $Q^2$-rich glass.  On the other hand, the
change in character with $K_e$ of the rigidity  transition is weak and
second order.  Finally, some chemical self-organization of the network
is allowed through an adaptative speciation.   As stress costs energy,
is  is natural  to   imagine that  the  glass   network will    try to
self-organize in the stressed  rigid phase to  decrease the energy  by
rewiring and reset  some weaker bonds such  as the ionic M-X ones that
will lead to a $Q^n$ specie recombination. In the present description,
this  means  that  at    a   given concentration   $x<x_{c(2)}$,   the
minimization of  the free energy can  be accomplished with  respect to
$K_e$.  The  equilibrium  constant $K_e$  minimizing ${\cal F}(x,K_e)$
provides  then  an  estimation  of  the   speciation  {\em via}   equ.
(\ref{1}) and the probability  of stressed rigid clusters.  This leads
to  a stress transition at  $x=x_{c(2)}=0.168$ and a free energy (dots
in the insert of Fig.
\ref{energy}) that is very close to the $K_e=0.1$ speciation model and to
sodium silicates \cite{Mysen}.
\par
In  summary, with   decreasing modifier concentration   $x$ there  are
different ways  for a  flexible  system to self-organize  in order  to
avoid stress, either  by nucleating weak stress  in small rings (EST),
or by producing
more flexible local structures ($Q^2$'s) that  balance the addition of
new constraints  arising from the   decrease of the  modifier content.
Both delay the  onset of  stressed  rigidity.   As  a conclusion,  we
provide   a prediction of the IP    for sodium seleniosilicates (M=Na,
X=Se) that have an EST fraction of $\eta=0.50$ in the base network former
$SiSe_2$
\cite{Ten}  and  an equilibrium   constant  $K_e=0.15$ \cite{Pradel1}.
According  to the  present  approach, one  therefore  expects a stress
transition    at  $x_{c(2)}=0.113$,     a   rigidity   transition   at
$x_{c(1)}=0.182$ and  a width for the Intermediate  Phase of  about $\Delta
x=0.069$.  Larger structural  correlations  will probably refine  this
picture and are under consideration.
\par
It is a pleasure to acknowledge ongoing discussions with P. Boolchand,
B. Goodman, M. Malki and  P. Simon. LPTMC is  Unité Mixte de Recherche
du Centre National de la Recherche Scientifique (CNRS) n. 7600.

\end{document}